# Regression Based Approach for Measurement of Current in Single-Phase Smart Energy Meter

Shashank Singh *Student Member, IEEE*, Arun S L, Selvan M P *Member, IEEE*
Hybrid Electrical Systems Laboratory
Department of Electrical and Electronics Engineering
National Institute of Technology Tiruchirappalli
{shashanksingh0110, slarun2010}@gmail.com, selvanmp@nitt.edu

*Abstract*—In the present scenario rate of utilization of electrical energy is continuously increasing, to preserve the non-renewable energy resources, efficient consumption of electrical energy is one of the most important needs of this era. Smart Energy Meter (SEM) plays a vital role in efficient use of electrical energy in consumer premises. Accuracy in the measurement of current is one of the key factors to be considered while measuring energy consumption. This paper introduces Arduino based low cost single-phase SEM and proposes a linear regression based technique for accurate measurement of current. The proposed current measurement approach gives accurate measurement for linear as well as highly non-linear loads. The SEM built in the laboratory has data transfer capability and also embedded with Demand Response (DR) feature. Block Rate Tariff structure used by Tamil Nadu Generation and Distribution Company is considered to incorporate DR that warns the consumers whenever their energy consumption reaches the boundary of a block.

## I. INTRODUCTION

Researchers have been continuously making efforts to increase the efficiency of power generation, transmission and distribution. Demand Side Management (DSM) is one of the modern-day approaches adopted by utilities to increase conservation and efficient consumption of energy. In DSM, it is expected that consumers will alter their energy usage pattern to reduce the electricity bill [1-3]. *Palensky* and *Dietrich* [4] have demonstrated impact of improved energy efficiency versus demand response while considering that customer driven activity might dominate in future.

SEM is one of the fundamental building blocks of automatic meter reading (AMR) and Energy Management Systems (EMS). EMS, based on AMR, have been proposed by many researchers. *Ashna et al.,* [5] have developed GSM based metering and billing system by using dedicated energy metering IC MCP3905A. *Rahman et al.,* [6] have proposed Arduino and GSM based SEM using energy meter IC BL6503 and concluded with the need of reliability and higher degree of satisfaction. *Bat-Erdene et al.,* [7] have extended the role of SEM to remote detection of illegal electricity usage with emphasis on installation of *gateway smart meter* and *terminal smart meter* for real time detection of illegal usage which reflects that scope of SEMs is not only limited to household energy monitoring but can be extended to many areas of power system. ZigBee based AMR system is proposed by *Prapasawad et. al,* [8] which calculates energy by using built-in functions. *Seong Ho Ju et. al.,* [9] initiated their literature stating that most of the energy policies and research projects seem unreal, inefficient and complex. They have proposed efficient home EMS using AMR and powerline communication. History, benefits, implementation and communication technology involved in meters and AMR are explained in review papers [10] and [11]. Development of Intelligent Residential Energy Management System (IREMS) [12] is proposed by *Arun and Selvan,* in which SEM is a part of.

The SEM consists of current sensor to sense the analog load current and feed the corresponding voltage to the analog to digital converter of the microcontroller used in it. Sensitivity of current sensor is defined as the ratio of sensor's output voltage to actual load current flowing through it. Significant change in the sensitivity of low cost hall effect current sensor (*ACS 712*) [13] is observed for same load current with different nature of load. Under such circumstances, it is not possible to assign a fixed value of sensitivity in program running inside a microcontroller to calculate the energy consumption. In this paper, A SEM embedded with regression based current measurement algorithm is developed as a solution of the observed problem for accurate current measurement using comparatively low cost sensors available in the market. Linear Regression is one of the important tool of machine learning. *Su et al.,* [14] have stated that linear regression aims to create linear relationship between two sets of variables and is used in the field of batch processing analysis, calibration of chemicals, machine learning and video surveillance. *De Matos et al.,* [15] have used regression models to explain distortion caused by non-linear loads.

This paper is structured in following way: In section II, voltage and current sensing units of SEM, features of developed SEM and metering algorithm are explained. Section III deals with proposed current measurement algorithm followed by experimental results in Section IV.

## II. SMART ENERGY METER OVERVIEW

### A. Smart Energy Meter Architecture

Electrical energy (*e*) is function of voltage (*v*), current (*i*) and time (*t*) expressed as given in (1) and accurate measurement of these quantities is prime objective of all energy meters.

$$e = \int_0^t (v.i)dt \qquad (1)$$

Analog voltage and current signals of the load circuit are sensed with the help of respective sensing units. However, the 10-bit unipolar Analog to Digital Converter (ADC) of Arduino microcontroller used in this work can only sample positive cycle of signals. Hence, a dc offset is given to the output of sensors to make Arduino capable of sampling both positive and negative cycles. A program that runs inside Arduino calculates rms value of voltage and current, active power, apparent power, power factor and energy from acquired samples. Calculated values are displayed on Liquid Crystal Display (LCD) and also transmitted through selected communication medium. Block diagram of proposed system and experimental setup are depicted in Fig. 1 and Fig. 2 respectively.

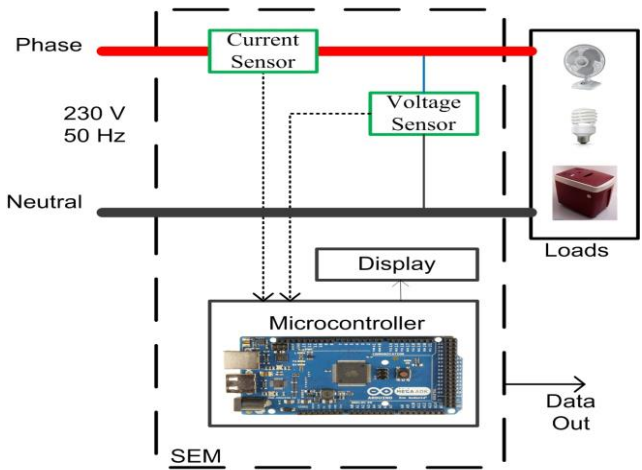

Fig. 1. Architecture of Proposed SEM

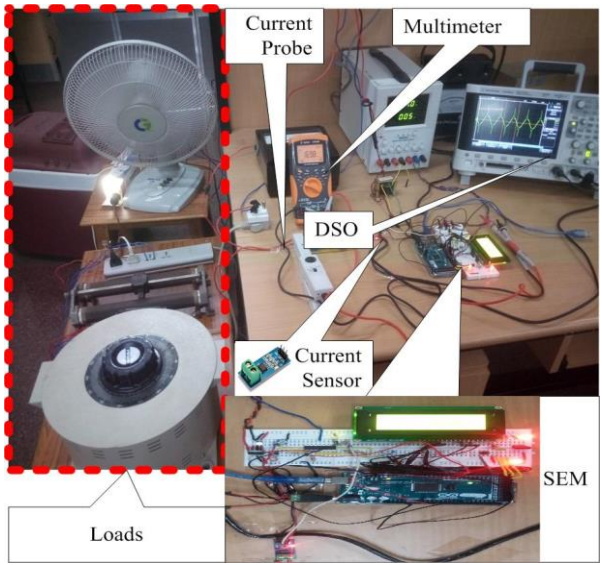

Fig. 2. Experimental Setup

### B. Determination of Electrical Parameters

Electrical quantities of interest are rms voltage, rms current, active power, reactive power, power factor and energy. RMS value of voltage and current are calculated using equation (2). Multiplication of rms value of voltage and current yields apparent power (S) and then active power (P) is calculated using equation (4) which is divided by S to obtain power factor (PF). Built in timers of Arduino are used to calculate precise time, which multiplied with P gives energy.

$$f(n) = \sqrt{\frac{1}{n}\left(\sum_{i=1}^{n} x_i^2\right)} \qquad (2)$$

$$S = V_{rms} * I_{rms} \qquad (3)$$

$$P = \frac{1}{n}\sum_{i=1}^{n} V_i * I_i \qquad (4)$$

$$PF = \frac{P}{S} \qquad (5)$$

where $n$ is the number of samples.

### C. Voltage Sensing Unit

Single-phase ac supply voltage is fed to a single-phase step down transformer, output of which is given dc offset and fed to operational amplifier based summer circuit. Potentiometer present at the output of summer circuit further steps down the voltage signal and feeds to Arduino as indicated in Fig. 3.

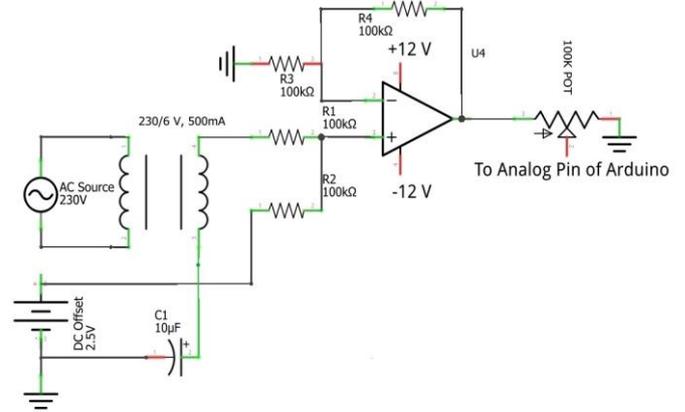

Fig. 3. Voltage Sensing Circuit

Analog read function is used in the program running in Arduino to sample the analog inputs, which takes 100 microseconds to read one sample and maps the input in the range of (0-1023) where 0 corresponds to 0 volt and 1023 to 4.99 volt. Samples containing only source signal are obtained and stored in an array after subtracting dc offset from input samples. Reconstruction of input signal scaled between 0-5 volt from stored samples has been done using Arduino serial plotter, which is shown in Fig. 4. This demonstrates the speed of Arduino ADC.

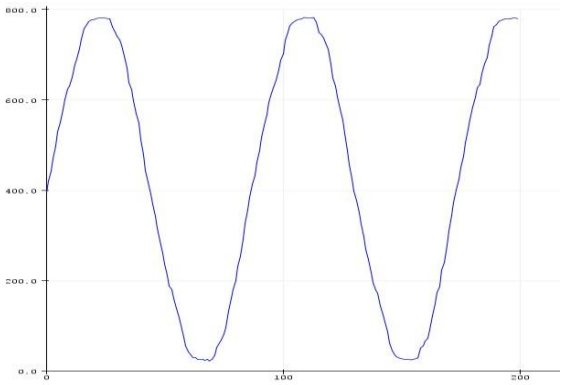

Fig. 4. Input voltage Scaled between 0-5 volt

### D. Current Sensing Unit

Hall effect current sensors are available with response time in the order of microseconds. For nominal current, up to 20 A, small footprint, low-profile SOIC8 packaged hall die is used [16], [17]. ACS 712-20-A current sensor is used as current transducer in this work that generates output voltage signal in the order of milli-volt having same wave shape as of load current, which is summed up with dc offset of $V_{cc}/2$. Connection diagram of current sensor is depicted in Fig. 5.

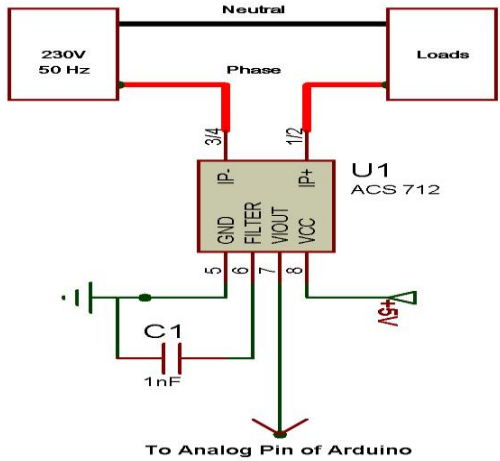

Fig. 5. ACS 712-20A as Current Transducer

### E. Feature 1: Automatic Meter Reading (AMR)

AMR is one of the early drivers of Smart Meters and with the application of microcontroller based metering technology it is easy to deploy energy meters which can send data to remote servers. Proposed SEM has capability of encoding data as shown in Fig. 6 and sending to desired location though the selected communication medium.

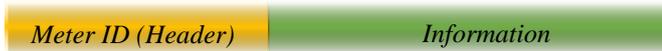

Fig. 6. Data Format

Header file contains identity of meter which helps in data storage and database management if multiple meters are sending data. Major benefit of automatic meter reading is reduction of cost and errors incurred by manual meter reading.

### F. Feature 2: Demand Response (DR)

Demand response depends upon energy usage pattern of a consumer and it can play significant role in efficient consumption of energy. Block rate tariff structure is considered to incorporate DR in which consumers are charged depending upon their category of consumption. SEM also generates warning signals if the consumer's consumption is about to reach maximum limit of consumption within a block. Structure of the block rate tariff considered in this work is shown in Table I.

TABLE I
IMPLEMENTED BLOCK RATE TARIFF STRUCTURE

| S. No | Units of Energy Consumption in a month | Price ($) |
|---|---|---|
| 1 | Less than X unit | a/unit |
| 2 | Between X and Y unit | b/unit (b > a) |
| 3 | More than Y unit | c/unit (c > b) |

### G. Algorithm

A computer program known as *Arduino Sketch* is compiled and loaded into microcontroller of Arduino board which does all the computational work for SEM. Pseudo-code of the program is given in Table II.

TABLE II
PSEUDO-CODE

**Begin**
*Timer* configured;
*GPIO* configured;
*Communication_Protocol* configured
*ADC* configured;
*Regression_Polynomial* Initialized;
Energy$_{old}$ = EEPROM
**while** True:
   *Acquire* voltage & current samples;
   *Subtract* dc offset from samples;
   *Calc* V$_{rms}$, I$_{rms}$ and P;
   *Obtain* S
   *Determine* PF
   **if** I$_{rms}$ != 0:
     Energy$_{new}$ = Energy$_{old}$ + P * ΔTime;
     EEPROM = Energy$_{new}$;
   **end of if**
   **else if** I$_{rms}$ = 0:
     Energy$_{new}$ = Energy$_{old}$;
   **end of else if**
   *Display* data on LCD
   *Send* data through communication channel
**end of while**
**end of program**

## III. CURRENT SENSING AND REGRESSION BASED APPROACH

Sensitivity of current sensor should be 100 mV/A according to datasheet but experimentally it was found that its sensitivity was neither the one as described nor varying with same factor. *Agilent 1146B* current probe along with *Agilent DSO-X-2014A* is used to measure reference value of current and *Agilent U1242B True RMS Multimeter* is used to measure sensor's output voltage in the range of milli-volt. The two types of loads used in the experiment are represented as Type-A load and Type-B load as detailed in Table III.

TABLE III
LOAD DESCRIPTION

| S. No | Types of Load | Detail |
|---|---|---|
| 1 | Type A | Auto-Transformer (1 Ph, 240 V, 28 A, 50/60 Hz), Rheostat (9 Ω, 10 A) |
| 2 | Type B | Type A + Compact Fluorescent Lamp (27W, 0.5PF), Table Fan (50W), Portable Fridge (230V, 0.53 A) |

Firstly, in the experimentation, output of current sensor with respect to load current is measured for type-A load then for type-B load in the range of 0-8 Ampere using above mentioned reference probes and a bar graph is plotted between sensor's output voltage and input load current as shown in Fig. 7.

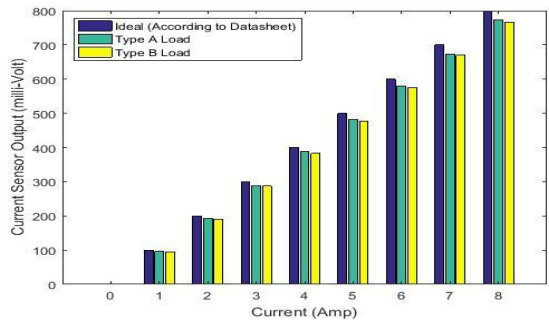

Fig. 7. Variation of Sensor's Output with respect to Load Current

It is clear from Fig. 7 that height of bars is not equal for same value of current for different nature of loads. Because of such difference among height of bars, *hard coded* value of sensitivity is replaced in this work with dynamic model using linear regression.

*Kantikoon et al.,* [18] have utilized multiple linear regression based approach for estimation of energy loss and *Cheikh et al.,* [19] have reported regression based predictive model having higher accuracy in prediction. In the present work, the proposed model is optimized in context of sensor's output to establish a relationship between response variable and independent variable by using training data. Training data are datasets containing load current as independent variable and sensor's output voltage as response variable. Output of current sensor is chosen as independent variable and current through load as response variable. Training datasets are measured using reference measuring instruments and analyzed using MATLAB curve-fitting toolbox. Plot obtained using *Polyfit* function is found as perfect fit hence 3$^{rd}$ order polynomial is preferred for ease of calculation inside microcontroller.

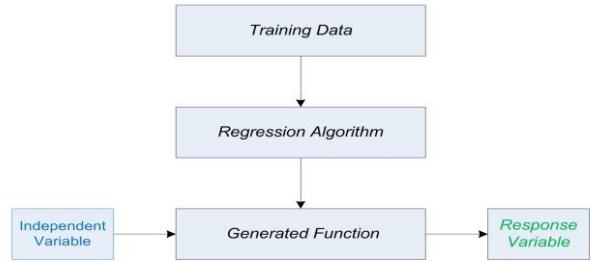

Fig. 8. Regression Model

Fig. 8 depicts steps which are involved in establishing relationship between response and independent variable and generated function from regression algorithm (*MATLAB polyfit function*) is expressed as (6).

$$f(x) = p_1 * x^3 + p_2 * x^2 + p_3 * x + p_4 \quad (6)$$

Where,
$p_1 = 0.9188 \quad p_2 = -1.406$
$p_3 = 10.86 \quad p_4 = -0.08648$
$x = \text{sensor's output}$
Load Current, $I_{rms} = f(x)$

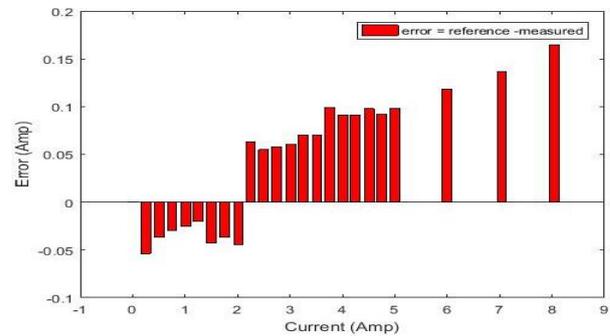

Fig. 9. Error in Current without Implementing Regression Approach

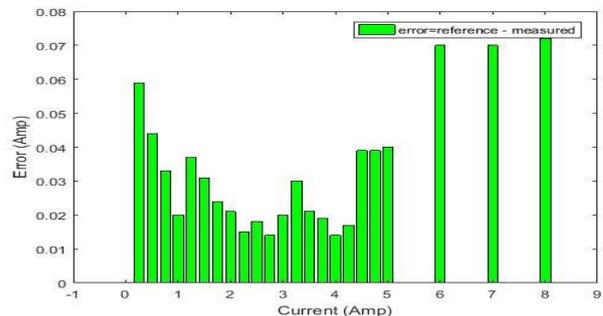

Fig. 10. Error in Current after Implementing Regression Approach

Proposed SEM is tested for current measurement at first with fixed value of sensitivity suggested by the datasheet of sensor and then at second with dynamic value of sensitivity by implementing generated regression polynomial. Current is increased up to a maximum value of 8 A by connecting various types of loads. The difference between DSO displayed value and SEM displayed value termed as error is illustrated in Fig. 9 and

Fig. 10 for the case without and with regression approach respectively. It is observed from the figures that maximum error has reduced from 150 mA to 70 mA by implementing regression approach. This experimental analysis also demonstrates significant improvement in the accuracy of current measurement by using low cost current sensor.

## IV. EXPERIMENTAL RESULTS

SEM is operated for 5 minutes to observe difference in calculation of energy for previously mentioned two cases. Energy is calculated in watt-hour and experimental data is detailed in Table IV.

TABLE IV
ENERGY CALCULATION

|  | Without Regression | With Regression | Difference |
|---|---|---|---|
| Energy Consumption (Wh) (5 minutes) | 31.61 | 32.1 | 0.49 |
| Predicted Energy Consumption (Wh) (31 days) | 282214.08 | 286588.8 | 4374.72 |

Proposed algorithm does not only improve accuracy in the measurement of current but also improves accuracy in energy measurement that has proportional relation with the cost incurred to the consumer.

## V. CONCLUSION

This paper presented behavior of current sensor for different types of loads. Regression based algorithm is proposed to improve accuracy in the measurement of current as well as energy. A low-cost SEM is presented which has demand response and data transfer capability. The experimental results have proved the validity of proposed algorithm and demonstrated significant improvement in accuracy of measurement and hence the cost incurred to the consumer.

## AKNOWLEDGEMENT

This project is supported and financed by Ministry of Electronics and Information Technology, Government of India.